\documentclass[a4paper]{jpconf}
\usepackage{microtype} 
\usepackage{supertabular}
\usepackage{array}
\newcolumntype{C}[1]{>{\centering\arraybackslash}m{#1}}
\newcolumntype{R}[1]{>{\raggedleft\arraybackslash}m{#1}}
\usepackage{amsmath}
\usepackage{amsfonts}
\usepackage{amssymb}
\usepackage{amscd}
\usepackage{cancel}
\usepackage{ntheorem}
\usepackage[american]{babel}
\usepackage[T1]{fontenc}
\usepackage{textcomp}
\usepackage[dvips]{rotating}
\usepackage{fancybox}
\usepackage{eurosym}
\usepackage[utf8]{inputenc}
\usepackage{epsfig}
\usepackage[pdftex]{color}
\usepackage{graphicx}
\usepackage{lmodern}
\usepackage{array}
\usepackage{hyperref}

\renewcommand{\d}[2]{\frac{d #1}{d #2}} % for derivatives
\newcommand{\ket}[1]{\left| #1 \right>} % for Dirac bras
\let\baraccent=\= % rename builtin command \= to \baraccent
\renewcommand{\=}[1]{\stackrel{#1}{=}} % for putting numbers above =

\renewcommand{\b}[1]{{\bf #1}}
\newcommand{\bmat}{\begin{pmatrix}}
\newcommand{\emat}{\end{pmatrix}}
\newcommand{\bdet}{\begin{dmatrix}}
\newcommand{\edet}{\end{dmatrix}}

\newcommand{\ra}{\rightarrow}

\newcommand{\tn}[1]{\textnormal{#1}}

\newcommand{\refeqn}[1]{Eq.~(\ref{#1})}
\newcommand{\reffig}[1]{Fig.~\ref{#1}}
\newcommand{\refsec}[1]{Sec.~\ref{#1}}

\renewcommand{\e}{\textnormal{e}}

\renewcommand{\i}{\mathrm{i}}
\renewcommand{\d}{\mathrm{d}}

\begin{document}
\title{Nonequilibrium Green's function approach to the pair distribution function of quantum many-body systems out of equilibrium}

\author{M~Bonitz$^1$, S~Hermanns$^1$, K~Kobusch$^1$, and K~Balzer$^2$}
\address{$^1$ Institut f\"ur Theoretische Physik und Astrophysik,
Universit\"at Kiel, D-24098 Kiel, Germany}
\address{$^2$ University of Hamburg, Max Planck Research Department for Structural Dynamics
Building 99 (CFEL), Luruper Chaussee 149, D-22761 Hamburg, Germany}
\ead{bonitz@theo-physik.uni-kiel.de}

\begin{abstract}
The pair distribution function (PDF) is a key quantity for the analysis of correlation effects of a quantum system both in equilibrium and far from equilibrium. We derive an expression for the PDF in terms of the single-particle Green's functions---the solutions of the Keldysh/Kadanoff-Baym equations in the two-time plane---for a one- or two-component system. The result includes initial correlations and generalizes previous density matrix expressions from single-time quantum kinetic theory. Explicit expressions for the PDF are obtained in second Born approximation.
\end{abstract} 
%\pacs{79.60.-i, 41.50.+h}

\section{Introduction}

In recent years the interest in correlated quantum many-body systems has increased steadily due to the observation of correlation effects such as liquid and crystal formation, superfluidity of Bose systems or bound states. Examples of systems include nuclear matter \cite{danielewicz84,garny09}, solid state systems \cite{haug96} and dense astrophysical, laboratory dusty plasmas or the quark gluon plasma, for a recent overview see \cite{bonitz12,kremp05}. A quantity sensitive to spatial correlations is the pair distribution function (PDF) $h(r)$---the probability to find an arbitrary pair of particles at a distance $r$.

Of particular current interest is the short-time behavior of correlated systems following external perturbation such as excitation by intense radiation, e.g., \cite{krausz09,krasovskii07}. In quantum systems such as atoms or solids this is often connected with rapid electron thermalization coupled to the dynamics of electronic correlations. Time and space-resolved measurement techniques detecting chemical reaction products, nuclear collision fragments or electrons and ions produced by laser ionization, e.g., \cite{becker08}, are becoming increasingly powerful. This brings the direct measurement of time-dependent pair correlations described by the nonequilibrium generalization of the PDF, $h(r,t)$, within reach and increases the need for theoretical and computational tools that are able to predict $h(r,t)$.

From the theory side PDFs are routinely computed for nonideal classical systems using, e.g., Monte Carlo or molecular dynamics simulations, e.g., \cite{bonitz12}. Extensions to quantum systems in equilibrium are available, e.g., in the frame of quantum Monte Carlo methods. However, there is still a high demand for accurate nonequilibrium simulations. Theoretical access to nonequilibrium PDFs is straightforwardly reached within density operator theory, e.g., \cite{bonitz96,kremp97,bonitz98}.
An alternative approach to quantum many-body systems out of equilibrium is based on nonequilibrium Green's functions (NEGF). Apart from exact time-dependent numerical calculations, e.g., \cite{bauch08}, which are limited to a few particles, numerical results for the two-time NEGF are expected to be the most accurate ones for strongly correlated quantum systems in nonequilibrium. This concerns, in particular, full two-time calculations, i.e. direct solutions of the Keldysh/Kadanoff-Baym equations (KBE) which, after the pioneering work of Danielewicz \cite{danielewicz84}, has now become routine  \cite{garny09,bonitz96j,bonitz97,kwong00,dahlen07,stan09,vonfriesen09}. They yield the single-particle Green's functions and all one-particle observables, including spectral function, as a function of time. A number of two-particle quantities can also be computed, taking advantage of the two-time structure of the equations, most importantly the interaction energy. However, in this approach the two-particle Green's function is eliminated by introduction of the selfenergy $\Sigma$ (the integration contour $\mathcal{C}$ and the arguments of the functions will be explained in \refsec{sec:bse}),
\begin{equation}
\label{eq:Selfenergy}
\pm \i\int_\mathcal{C}\mathrm{d}2\,V(1-2)G(12,1'2^+)=\int_\mathcal{C}\mathrm{d}2\,\Sigma(1,2)G(2,1')\,.
\end{equation} 
This concept of the selfenergy has proven extremely successful for reducing the many-particle problem to an effective single-particle one (quasiparticle picture) which forms the basis of Green's function theory and the Feynman diagram technique. At the same time, the direct access to the pair distribution function---which contains important additional information beyond the one carried by the single-particle Green's function---is lost. Due to this reason, so far, no NEGF results for the full pair distribution function have been obtained which go beyond the computation of double occupancies, accessible from \refeqn{eq:Selfenergy} \cite{vonfriesen11}. 

In this paper we solve this problem. Starting from the Bethe-Salpeter equation for the two-particle Green's function, we demonstrate in \refsec{sec:bse} how, from a given time-dependent solution for the single-particle Green's function in a chosen approximation for the selfenergy $\Sigma$, the PDF in the same approximation can be reconstructed. In this paper, we choose, as an example, the second order Born selfenergy approximation $\Sigma^\tn{2B}$ which includes the interaction up to second order ($\Sigma^\tn{2B}\propto V^2$). Our results include the standard PDF $h(r)$ in equilibrium, and its extension $h({\bf r}_1,{\bf r}_2)$ to inhomogeneous systems and to multicomponent systems where the PDF becomes a matrix $h^\tn{ab}({\bf r}_1,{\bf r}_2)$, $a,b$ labeling the particle species. Further, we obtain results for an arbitrary nonequilibrium system where the PDF becomes time-dependent, $h^\tn{ab} \rightarrow h^\tn{ab}(t)$. Finally, the results are directly generalized to two-time 
PDF's, $h^\tn{ab}({\bf r}_1,{\bf r}_2) \rightarrow h^\tn{ab}({\bf r}_1,{\bf r}_2,t)$. Our results selfconsistently include the influence of a correlated initial state, and they describe the decay of initial correlations which is particularly important at the initial stage of the relaxation. This 
problem was studied in detail for the single-time PDF within density operator theory \cite{bonitz96,kremp97,bonitz98,bonitz96p}, and these results are contained in our theory as a special case.

In \refsec{sec:bilayer} we discuss, as an illustration, the possible application of the PDF reconstruction algorithm to a two-component system of electrons and holes in a bilayer structure. This system has attracted substantial interest in recent years because it exhibits strong electron-electron and hole-hole correlations leading to spatial ordering and, at the same time, electron-hole bound states (excitons). There have been detailed investigations of the phase diagram \cite{depalo02,filinov03,hartmann05,schleede12}, of exciton and hole crystallization \cite{filinov03,filinov09, ludwig07}, of collective excitations \cite{ludwig08,kalman07} and of exciton Bose condensation and superfluidity, e.g., \cite{filinov09,boening11}. We conclude with a summary and discussion in \refsec{sec:discussion}.
\section{Bethe-Salpeter equation}\label{sec:bse}
In nonequilibrium Green's functions theory, the PDF of particle species ``a'' and ``b'' is obtained from the two-particle correlation function \cite{kremp05}
taken at four equal times $t_1=t_2=t'_1=t'_2=t$,
\begin{eqnarray}\label{eq:hab}
 h^{\tn{ab}}({\bf r}_1,{\bf r}_2,{\bf r}_1,{\bf r}'_2,t) &=& 
\i^2 g^{\tn{ab},<}(1,2,1',2')
\\
&=& 
\langle \Psi_\tn{a}^{\dagger}(1')\Psi_\tn{b}^{\dagger}(2')\Psi_\tn{b}(2)\Psi_\tn{a}(1)\rangle\,,
\nonumber
\end{eqnarray}
where $\Psi_\tn{a}^{\dagger}$ and $\Psi_\tn{a}$ are fermionic or bosonic creation and annihiliation operators, and 
we introduced the short-hand notation $1={\bf r}_1, s_{z1}, t_1$. To determine the two-particle correlation function $g^{\tn{ab},<}$ in nonequilibrium we start from the more general two-particle Green's function $g^{\tn{ab}}$ on the Schwinger-Keldysh contour ${\cal C}$ which consists of a Hartree-Fock and a correlation part
\begin{eqnarray}
\label{eq:g_ab}
g^{\tn{ab}}(12,1'2')&=&g_\tn{HF}^{\tn{ab}}(12,1'2')+g_\tn{corr}^{\tn{ab}}(12,1'2')\,, \\
g_\tn{HF}^{\tn{ab}}(12,1'2')&=& g^\tn{a}(1,1') g^\tn{b}(2,2')  \pm \delta_{\tn{ab}} g^\tn{a}(1,2') g^\tn{b}(2,1')\,, 
\end{eqnarray}
where we allow for four different time arguments, $t_1,t_2,t_1',t_2'$, all located on the contour $\mathcal{C}$. The correlation part, $g_\tn{corr}^{\tn{ab}}$, obeys the Bethe-Salpeter equation, e.g., \cite{bornath99},
\begin{equation}
\label{eq:bse}
g_\tn{corr}^{\tn{ab}}(12,1'2') = \i \int_{\cal C} \d {\bar 1}\d {\bar 2}\d {\tilde 1}\d {\tilde 2} \,
g^\tn{a}(1,{\bar 1}) g^\tn{b}(2,{\bar 2})
K^\tn{ab}({\bar 1} {\bar 2}, {\tilde 1}{\tilde 2})\,g^\tn{ab}({\tilde 1}{\tilde 2},1'2')\,,
\end{equation}
which presents a formal closure of the second equation of the Martin-Schwinger hierarchy for the two-particle Green's function. Compared to Ref.~\cite{bornath99} we also restored the exchange term where plus (minus) refers to bosons (fermions). In \refeqn{eq:bse} the function $K^\tn{ab}$ is a general dynamic interaction (for details see Ref.~\cite{kremp05}), which will be simplified in the following.

1.) The first simplification is to introduce the screened ladder approximation, $K^\tn{ab}({\bar 1} {\bar 2}, {\tilde 1}{\tilde 2}) \rightarrow 
V^\tn{ab}({\bar 1} {\bar 2}) \delta({\bar 1}-{\tilde 1})\delta({\bar 2}-{\tilde 2})$. Suppressing for a moment the space integration and spin summation (they are implied by the repeated arguments ${\bar 1}$ and ${\bar 2}$ under the integral),
we obtain from \refeqn{eq:bse}
\begin{eqnarray}
g_\tn{corr}^\tn{ab}(1 2,1' 2') = \i \int_{\cal C} \d {\bar t}_1 \d {\bar t}_2 \,
g^\tn{a}(1,{\bar 1}) g^\tn{b}(2,{\bar 2}) 
V^\tn{ab}({\bar 1} {\bar 2})\,g^\tn{ab}({\bar 1} {\bar 2},1'2')\,.
\quad
\label{eq:bse_vs} 
\end{eqnarray}
We mention that general nonequilibrium initial correlations can be included by the proper definition of the contour $\mathcal{C}$ or accounted for via additional contributions to the single particle selfenergy, e.g., \cite{bonitz97,semkat99,semkat00,semkat03}. Equation (\ref{eq:bse_vs}) contains strong coupling and dynamical screening effects leading to a complicated integro-differential equation. 

2.) Our next simplification is to neglect dynamic effects in the potential (dynamical screening, related to the GW approximation) which leads to the replacement $V^\tn{ab}({\bar 1} {\bar 2}) \rightarrow V^\tn{ab}({\bar r}_{12})\delta({\bar t}_1-{\bar t}_2)$, where ${\bar r}_{12} = |{\bf r}_1-{\bf r}_2|$,
\begin{eqnarray}
g_\tn{corr}^\tn{ab}(1 2,1' 2') = \i \int_{\cal C} \d {\bar t} \,
g^\tn{a}(1,{\bar 1}) g^\tn{b}(2,{\bar 2}) 
V^\tn{ab}({\bar r}_{12})\,g^\tn{ab}({\bar 1} {\bar 2},1'2')\,,
\quad
\label{eq:bse_static} 
\end{eqnarray}
and under the integral ${\bar t}_1={\bar t}_2={\bar t}$. This equation corresponds to a static T-matrix approximation for the two-particle Green's function and the pair correlations, i.e. to a complete summation of the Born series. 

3.) Since our goal is to reconstruct the pair distributions from a solution of the two-time KBE for the single-particle Green's functions in second order Born approximation, we limit ourselves to the first iteration of the integral equation (\ref{eq:bse_static}), i.e. we replace, under the integral, the two-particle Green's function by the Hartree-Fock approximation:

\begin{equation}
g^\tn{ab}(1 2,1' 2') = \i \int_{\cal C} \d {\bar t} \,
g^\tn{a}(1,{\bar 1}) g^\tn{b}(2,{\bar 2}) 
V^\tn{ab}({\bar r}_{12})g_\tn{HF}^\tn{ab}(1 2,1' 2')\,,\quad \mbox {with}\quad {\bar t}_1={\bar t}_2={\bar t}\,.
\label{eq:bse_static_born} 
\end{equation}
We can bring this equation to a more compact form by introducing the following 
new functions
\begin{eqnarray}
\label{eq:G0ab}
G^\tn{ab}_0(1 2,1' 2') & = & g^\tn{a}(1,1') g^\tn{b}(2,2'),
\\[2ex]
\Sigma^\tn{ab}_0(1 2,1' 2') & = & V^\tn{ab}(r_{12})\left\{ 
G^\tn{ab}_0(1 2,1' 2') \pm \delta_\tn{ab} G^\tn{ab}_0(1 2,2' 1')
\right\}\,,
\label{eq:Sigma0ab}
\end{eqnarray}
where $G^\tn{ab}_0$ is the Hartree approximation for the two-particle Green's function ($g_\tn{HF}^\tn{ab}$ without exchange terms) and $\Sigma^\tn{ab}_0$ the first-order two-particle selfenergy. With these definitions \refeqn{eq:bse_static_born} becomes
\begin{equation}
g^{ab}_\tn{corr}(1 2,1' 2') =  \i \int_{\cal C} \d {\bar t} \,
G^\tn{ab}_0(1 2,{\bar 1} {\bar 2})\Sigma^\tn{ab}_0({\bar 1} {\bar 2},1' 2')\,,\quad\mbox {with}\quad {\bar t}_1={\bar t}_2={\bar t}\,.
\label{eq:bse_static_born_short} 
\end{equation}
We underline that all expressions, so far, are written on the Schwinger-Keldysh contour, i.e. 
all single-particle Green's functions are $3\times 3$ matrices whereas the two-particle functions contain $3^4$ Keldysh matrix elements.
To obtain the nonequilibrium pair correlation function we now have to extract from \refeqn{eq:bse_static_born_short} functions depending on the real physical time. In particular, according to \refeqn{eq:hab}, we need only the two-particle correlation function $g^\tn{ab,<}$ which we consider in the following.

%-----------------------------
\section{Reduction of the time structure of the two-particle correlation function $g^\tn{ab,<}$}
In order to compute the pair distribution function, \refeqn{eq:hab}, the two-particle correlation function $g^\tn{ab,<}$ at four equal times is needed. To this end, we first determine the equation of motion for the two-time two-particle correlation function $g^\tn{ab,<}(t,t')$ and then specialize to the one-time two-particle correlation function $g^\tn{ab,<}(t)$.     
\subsection{Two-time two-particle correlation function $g^\tn{ab,<}(t,t')$}
The problem of the large number of Keldysh matrix elements of $g^\tn{ab}$ can be simplified drastically in the static Born approximation. As a first step in simplifying the time dependencies we specialize to functions with pairwise equal time arguments, $t_1=t_2=t$, and $t'_1=t'_2=t'$. Then it is clear from \refeqn{eq:bse_static_born_short} that the functions $G_0^\tn{ab}$ and $\Sigma_0^\tn{ab}$ depend only on two times in the following way
\begin{eqnarray}
G^\tn{ab}_0({\bf r}_1{\bf r}_2;{\bf r}'_1{\bf r}'_2;tt') & = & 
g^\tn{a}({\bf r}_1t;{\bf r}'_1t') g^\tn{b}({\bf r}_2t;{\bf r}'_2t')\,,
\\[2ex]
\Sigma^\tn{ab}_0({\bf r}_1{\bf r}_2;{\bf r}'_1{\bf r}'_2;tt') & = & V^\tn{ab}(r_{12})
\big\{G^\tn{ab}_0({\bf r}_1{\bf r}_2;{\bf r}'_1{\bf r}'_2;tt')
\pm \delta_\tn{ab} G^\tn{ab}_0({\bf r}_1{\bf r}_2;{\bf r}'_2{\bf r}'_1;tt')
\big\}\,,
\end{eqnarray}
where we also restored the coordinate arguments (spin variables are not written explicitly). Then \refeqn{eq:bse_static_born_short} turns into
\begin{eqnarray}
g^\tn{ab}_\tn{corr}({\bf r}_1{\bf r}_2;{\bf r}'_1{\bf r}'_2;tt') =  
 \i \int_{\cal C} \d {\bar t} \,
G^\tn{ab}_0({\bf r}_1{\bf r}_2;{\bar {\bf r}}_1 {\bar {\bf r}}_2,t{\bar t})
\Sigma^\tn{ab}_0({\bar {\bf r}}_1 {\bar {\bf r}}_2;{\bf r}'_1{\bf r}'_2;{\bar t} t')\,.
\label{eq:bse_static_born_short_1t} 
\end{eqnarray}
With this result, \refeqn{eq:g_ab} for $g^\tn{ab}$ can be rewritten as 
\begin{equation}
\label{eq:gabtt'}
g^{\tn{ab}}({\bf r}_1{\bf r}_2;{\bf r}'_1{\bf r}'_2;tt')=g_\tn{HF}^{\tn{ab}}({\bf r}_1{\bf r}_2;{\bf r}'_1{\bf r}'_2;tt')+g_\tn{corr}^{\tn{ab}}({\bf r}_1{\bf r}_2;{\bf r}'_1{\bf r}'_2;tt')\,,
\end{equation}
where $g^\tn{ab}_\tn{corr}$ is completed by the Hartree-Fock contribution which, in the new notation, acquires the form
\begin{eqnarray}
g^\tn{ab}_\tn{HF}({\bf r}_1{\bf r}_2;{\bf r}'_1{\bf r}'_2;tt')=G^\tn{ab}_0({\bf r}_1{\bf r}_2;{\bf r}'_1{\bf r}'_2;tt') \pm \delta_\tn{ab} G^\tn{ab}_0({\bf r}_1{\bf r}_2;{\bf r}'_2{\bf r}'_1;tt')\,.
\label{eq:gabHF} 
\end{eqnarray}
According to \refeqn{eq:hab}, $g^{\tn{ab}}(tt')$ reads in terms of creation and annihilation operators (suppressing spatial arguments),
\begin{equation}
g^{\tn{ab}}(tt')=\i^2\langle \Psi_\tn{a}^{\dagger}(t')\Psi_\tn{b}^{\dagger}(t')\Psi_\tn{b}(t)\Psi_\tn{a}(t)\rangle\,.
\end{equation}
It is important to notice that this structure is different from a generalized density-density correlation function
\begin{equation}
c_\tn{n-n}^\tn{ab}(tt')=\i^2\langle n_\tn{b}(t')n_\tn{a}^{\dagger}(t)\rangle=\i^2\langle \Psi_\tn{b}^{\dagger}(t')\Psi_\tn{b}(t')\Psi_\tn{a}^\dagger(t)\Psi_\tn{a}(t)\rangle\neq g^{\tn{ab}}(tt')\,,
\end{equation}
since the order of the operators, in general, cannot be interchanged. 

Now, to compute the pair distribution function from \refeqn{eq:gabtt'}, we have to extract $g^{\tn{ab},<}$ from the Keldysh matrix and, in particular, the ``$<$'' component from the contour integral in \refeqn{eq:bse_static_born_short_1t} . The solution of this problem is well known for one-particle functions, and for two-particle functions it has been solved in Ref.~\cite{bornath99}. However, the latter results are not needed here. Indeed, although we are dealing with two-particle functions $G_0^\tn{ab}$ and $\Sigma_0^\tn{ab}$, in the present approximation, they have the same time-dependence (they depend just on $t,t'$) as occurs in the case of single-particle functions in the integral term of the Keldysh 
Kadanoff-Baym equations. As a consequence, both these functions are $3\times 3$ Keldysh matrices and, therefore, also 
the two-particle Green's function $g^\tn{ab}$, \refeqn{eq:bse_static_born_short_1t}, is a $3\times 3$ Keldysh matrix. 

The Keldysh components are classified by location of the two time-arguments on the contour $\mathcal{C}$. They include correlation functions, ``$>,<$'' with two real arguments, ``$\lceil,\rceil$'' having one real and one imaginary time argument and the Matsubara component, $\alpha=\tn{M}$, where both arguments lie on the imaginary track. Below we will need only the ``$<$'' component which determines the time-dependent nonequilibrium PDF and the Matsubara component which yields the PDF in thermodynamic equilibrium. We start the analysis with the former and consider the latter in \refsec{sec:eqpdf}.
  
Each of the nine components  $G_0^{\tn{ab},\alpha}$ of the matrix $G_0^\tn{ab}$ is of the product form 
\begin{eqnarray}
\label{eq:g0abalpha}
G^{\tn{ab},\alpha}_0({\bf r}_1{\bf r}_2;{\bf r}'_1{\bf r}'_2;tt') & = & 
g^{\tn{a},\alpha}({\bf r}_1t;{\bf r}'_1t') g^{\tn{b},\alpha}({\bf r}_2t;{\bf r}'_2t')\,,
\end{eqnarray}
which allows for an easy matrix multiplication of $G_0^\tn{ab}$ and $\Sigma_0^\tn{ab}$ under the integral in \refeqn{eq:bse_static_born_short_1t}. Since each component $\alpha$ has the same dependence on the coordinates as in \refeqn{eq:bse_static_born_short_1t}, we will suppress the space arguments and integrations in the remainder of this section.

The result for the ``$<$'' component is well known (Langreth rules) and consists of an initial correlation and a collision term 
\begin{equation}
\label{eq:gabcorrless}
g^{\tn{ab},<}_\tn{corr}(t,t') = g^{\tn{ab},<}_\tn{IC}(t,t') +g^{\tn{ab},<}_\tn{col}(t,t')\,,
\end{equation}
which arise, respectively, from the imaginary and real part of the contour $\mathcal{C}$,
\begin{eqnarray}
\label{eq:gablessIC}
g^{ab,<}_\tn{IC}(t,t')&=& -\i\int_0^\beta\d\tau\,G^{\tn{ab},\rceil}_0(t,\tau)\Sigma^{\tn{ab},\lceil}_0(\tau,t')\,, \\
g^{ab,<}_\tn{col}(t,t')&=&\int_0^\infty\d\bar{t}\left\{G^{\tn{ab},<}_0(t,\bar{t})\Sigma^{\tn{ab,A}}_0(\bar t,t')+G^{\tn{ab},R}_0(t,\bar{t})\Sigma^{\tn{ab,<}}_0(\bar t,t')\right\}\,.
\label{eq:gablesscol}
\end{eqnarray}
Here $G^{\tn{ab},<}_0$ denotes the product of two single-particle correlation functions $g^{\tn{a},<},g^{\tn{b},<}$, both having two real time arguments, whereas the components $\rceil$ and $\lceil$ describe products of functions depending on one real and one complex ($\tau$) time argument describing the propagation of initial correlations, for a detailed discussion, see, e.g., Refs.~\cite{bonitz98,stan09}. Further, we introduced retarded and advanced functions defined by
\begin{equation}
\label{eq:spec_id}
a^\tn{R/A}(t,t')=\pm\Theta\left[\pm(t-t')\right]\left\{a^>(t,t')-a^<(t,t')\right\}\,.
\end{equation}
\subsection{One-time two-particle correlation function $g^\tn{ab,<}(t)$}
We now perform the final step on the way to the nonequilibrium PDF---taking the time-diagonal limit, $t=t'$, in the above equations for $g^{\tn{ab},<}$ which has the general structure
\begin{equation}
\label{eq:gablesst}
g^{\tn{ab},<}(t)=g_\tn{HF}^{\tn{ab},<}(t)+g_\tn{IC}^{\tn{ab},<}(t)+g_\tn{col}^{\tn{ab},<}(t)\,.
\end{equation}
Expressing the time-diagonal part of the single-particle correlation function by the single-particle density matrix $\rho^\tn{a}(t)=\pm \i\, g^\tn{a}(t,t)$, where the upper (lower) sign refers to bosons (fermions), we obtain for the two-particle correlation function on the time-diagonal
\begin{equation}
G_0^{\tn{ab},<}(t,t)=-\rho^\tn{a}(t)\rho^\tn{b}(t)\,.
\end{equation}
With this relation and \refeqn{eq:gabHF} we first obtain the nonequilibrium Hartree-Fock pair distribution function
\begin{equation}
\label{eq:gabHFt}
g^{\tn{ab},<}_\tn{HF}({\bf r}_1{\bf r}_2;{\bf r}'_1{\bf r}'_2;t)=-\rho^\tn{a}({\bf r}_1{\bf r}'_1,t)\rho^\tn{b}({\bf r}_2{\bf r}'_2,t) \mp \delta_\tn{ab} \rho^\tn{a}({\bf r}_1{\bf r}'_2,t)\rho^\tn{b}({\bf r}_2{\bf r}'_1,t)\,.
\end{equation}
The second contribution to \refeqn{eq:gablesst} due to the initial correlations follows from \refeqn{eq:gablessIC} and is given by
\begin{eqnarray}
\label{eq:gablessICt}
g^{\tn{ab},<}_\tn{IC}({\bf r}_1{\bf r}_2;{\bf r}'_1{\bf r}'_2;t) &=
-\i&\int\limits^{\beta}_0 \d\tau \int \mathrm{d}^3{\bar r}_1 \mathrm{d}^3{\bar r}_2 V^\tn{ab}({\bar r}_{12}) g^{\tn{a},\rceil}({\bf r}_1t;{\bar {\bf r}}_1\tau) g^{\tn{b},\rceil}({\bf r}_2t;{\bar {\bf r}}_2\tau) 
\\\nonumber
&&\times\bigg\{
g^{\tn{a},\lceil}({\bar {\bf r}}_1\tau,{\bf r}'_1,t) g^{\tn{b},\lceil}({\bar {\bf r}}_2\tau,{\bf r}'_2;t)
\pm \delta_\tn{ab}
g^{\tn{a},\lceil}({\bar {\bf r}}_1\tau,{\bf r}'_2,t) g^{\tn{b},\lceil}({\bar {\bf r}}_2\tau,{\bf r}'_1;t)
\bigg\}\,.
\end{eqnarray}
Finally, the scattering contribution (\ref{eq:gablesscol}) contains two real-time integrals extending to $t$ and $t'$, respectively. Since now both times are equal, a partial cancellation is possible. Indeed, taking advantage of relation (\ref{eq:spec_id}) allows to cancel all contributions with products of four ``$>$'' or four ``$<$'' functions, and we obtain 
\begin{eqnarray}
\label{eq:gablesscolt}
g_\tn{col}^{\tn{ab},<}({\bf r}_1{\bf r}_2;{\bf r}'_1{\bf r}'_2;t) &=
\i&\int\limits^{t}_0 \d\bar t \int \d^3{\bar r}_1 \d^3{\bar r}_2 V^\tn{ab}({\bar r}_{12})
\bigg\{
g^{\tn{a},>}({\bf r}_1t;{\bar {\bf r}}_1 {\bar t}) g^{\tn{b},>}({\bf r}_2t;{\bar {\bf r}}_2 {\bar t}) 
\\\nonumber
&&\times\big[
g^{\tn{a},<}({\bar {\bf r}}_1{\bar t},{\bf r}'_1,t) g^{\tn{b},<}({\bar {\bf r}}_2 {\bar t},{\bf r}'_2;t)
\pm \delta_\tn{ab}
g^{\tn{a},<}({\bar {\bf r}}_1{\bar t},{\bf r}'_2,t) g^{\tn{b},<}({\bar {\bf r}}_2 {\bar t},{\bf r}'_1;t)
\big]\nonumber\\
&&- (> \leftrightarrow <)
\bigg\}\,.\nonumber
\end{eqnarray}
This expression is readily understood: the first term describes the correlation build-up due to scattering of a particle pair $a,b$ out of state $\ket{\bar {\bf r}_1}\ket{\bar {\bf r}_2}$ into state $\ket{{\bf r}_1}\ket{{\bf r}_2}$, whereas the second contribution which is obtained by interchanging functions with the indices ``$<$'' and ``$>$'' describes the opposite process. The integral has the familiar non-Markovian structure (``memory'') indicating that scattering processes from all times prior to the current one do contribute although their weight decreases since the single-particle Green's function decay with increasing difference of their time arguments.
Expression (\ref{eq:gablesscolt}) has exactly the same structure as known from density operator theory \cite{bonitz98}. The binary 
density operator in Born approximation is recovered if the two-time correlation functions are eliminated using the generalized Kadanoff-Baym ansatz \cite{lipavsky86}, see also Ref.~\cite{hermanns12}. The present result is more general because this reconstruction ansatz, which is only valid approximately and becomes increasingly inaccurate with increasing correlation effects, is avoided.

%-----------------------------
\section{Nonequilibrium pair distribution function}
\label{sec:neqpdf}
The nonequilibrium pair distribution function $h^\tn{ab}(t)$ follows immediately from the two-particle correlation function according to \refeqn{eq:hab} and the results (\ref{eq:gablesst}), (\ref{eq:gabHFt}), (\ref{eq:gablessICt}) and (\ref{eq:gablesscolt}). Explicit results depend on the choice of a basis. We start from the coordinate representation.
\subsection{Coordinate representation}
\label{sec:NeqPDFcoord}
To obtain the standard pair distribution function which is defined in configuration space we set ${\bf r}_1={\bf r}'_1$
and ${\bf r}_2={\bf r}'_2$ in all expressions 
\begin{equation}
\label{eq:habcoord}
\begin{split}
 h^\tn{ab}({\bf r}_1,{\bf r}_2,t) =
&\rho^\tn{a}({\bf r}_1{\bf r}_1,t) \rho^\tn{b}({\bf r}_2{\bf r}_2,t)
\pm  \delta_\tn{ab}  \rho^\tn{a}({\bf r}_1 {\bf r}_2,t) \rho^\tn{b}({\bf r}_2 {\bf r}_1,t)\\
&-g^{\tn{ab},<}_\tn{IC}({\bf r}_1{\bf r}_2;{\bf r}_1{\bf r}_2;t) 
-g^{\tn{ab},<}_\tn{col}({\bf r}_1{\bf r}_2;{\bf r}_1{\bf r}_2;t)\,.
\end{split}
\end{equation}
In a spatially inhomogeneous system it is advantageous to introduce center of mass and relative coordinates, 
${\bf R}=({\bf r}_1+{\bf r}_2)/2$ and ${\bf r}={\bf r}_1-{\bf r}_2$, leading to the replacements 
${\bf r}_{1,2}={\bf R}\pm {\bf r}/2$, on the right hand side. Then $h^\tn{ab}({\bf R},{\bf r})$ is understood 
as the local probability to find a particle pair at distance ${\bf r}$ around space point ${\bf R}$.

If spatial inhomogeneities are not relevant or not of interest, a space integration leads to the space averaged nonequilibrium pair distribution
\begin{equation}
h^\tn{ab}(\b{r},t)=\int\d^dR\,h^\tn{ab}(\b{R},\b{r},t)\,,
\end{equation}  
where $d$ is the dimensionality of the system. Note that $h^\tn{ab}$ is normalized to the total number of particles according to
\begin{equation}
\label{eq:pdf_int}
\int \d^d h^\tn{ab}(\b{r},t)=N_\tn{a}N_\tn{b}\,.
\end{equation} 
Finally, in some cases just the dependence on the magnitude of the pair separation is of interest which is obtained by an angle integration. The result is called radial distribution function, $h^\tn{ab}(r,t)$, and follows from \refeqn{eq:pdf_int} according to (we consider a two-dimensional system and change to polar coordinates)
\begin{equation}
\label{eq:habrt}
h^\tn{ab}(r,t)=\int_0^{2\pi}\d\phi\, r h^\tn{ab}(r,\phi,t)\,,
\end{equation}
with the normalization
\begin{equation}
\int_0^\infty\d r h^\tn{ab}(r,t)=N_\tn{a}N_\tn{b}\,,
\end{equation} 
and similarly in 3D.
\subsection{Arbitrary basis}
\label{sec:NeqPDFbasis}
Let us now consider the case of an arbitrary orthonormal stationary basis $\left\{\phi_i(\b{r})\right\}$ which is of relevance, in particular, for spatially inhomogenenous systems. Then the single-particle Green's functions become matrices according to
\begin{equation}
g^{\tn{b},\alpha}(1,1')=\sum_{ij}\phi_i(\b{r}_1)g_{ij}^{\tn{b},\alpha}(t_1,t_1')\phi_i^*(\b{r}_1')\,,
\end{equation}
which holds for any Keldysh component ``$\alpha$''. Analogously, the two-particle single-time correlation function is represented by a four-dimensional matrix
\begin{equation}
\label{eq:gablessbasis}
g^{\tn{ab},<}(\b{r}_1\b{r}_2;\b{r}_1'\b{r}_2';t)=\sum_{ijkl}\phi_i(\b{r}_1)\phi_j(\b{r}_2)g_{ijkl}^{\tn{ab},<}(t)\phi^*_k(\b{r}_1')\phi^*_l(\b{r}_2')\,.
\end{equation}
To compute the nonequilibrium PDF (\ref{eq:gablesst}) we have to expand the Hartree-Fock, the initial correlation and the scattering part in this basis. For the Hartree-Fock contribution we obtain, in analogy to (\ref{eq:gablessbasis}),
\begin{eqnarray}
\label{eq:gablessHFtbasis}
g_\tn{HF}^{\tn{ab},<}(\b{r}_1\b{r}_2;\b{r}_1'\b{r}_2';t)&=&\sum_{ijkl}\phi_i(\b{r}_1)\phi_j(\b{r}_2)g_{\tn{HF},ijkl}^{\tn{ab},<}(t)\phi^*_k(\b{r}_1')\phi^*_l(\b{r}_2')\,,\\
g_{\tn{HF},ijkl}^{\tn{ab},<}(t)&=&g_{ik}^{\tn{a},<}(t)g_{jl}^{\tn{b},<}(t)\pm\delta_\tn{ab}g_{il}^{\tn{a},<}(t)g_{jk}^{\tn{b},<}(t)\,.
\end{eqnarray}
Similarly, we obtain from \refeqn{eq:gablessICt} for the initial correlation contribution
\begin{eqnarray}
\label{eq:gablessICtbasis}
g_\tn{IC}^{\tn{ab},<}(\b{r}_1\b{r}_2;\b{r}_1'\b{r}_2';t)&=&\sum_{ijkl}\phi_i(\b{r}_1)\phi_j(\b{r}_2)g_{\tn{IC},ijkl}^{\tn{ab},<}(t)\phi^*_k(\b{r}_1')\phi^*_l(\b{r}_2')\,,\\
g_{\tn{IC},ijkl}^{\tn{ab},<}(t)&=&\int_0^\beta \d\tau \sum_{mnrs} V^\tn{ab}_{mnrs}\,
  g^{\tn{a},\rceil}_{m,i} (t,\tau)g^{\tn{b},\rceil}_{n,j} (t,\tau)\, \times \\
&&\left\{
  g^{\tn{a},\lceil}_{r,k} (\tau,t)g^{\tn{b},\lceil}_{s,l} (\tau,t)
  \pm \delta_\tn{ab}\, g^{\tn{a},\lceil}_{r,l} (\tau,t)g^{\tn{b},\lceil}_{s,k} (\tau,t)
  \right\}\nonumber\,,
\end{eqnarray}
and from \refeqn{eq:gablesscolt}
\begin{eqnarray}
\label{eq:gablesscoltbasis}
g_\tn{col}^{\tn{ab},<}(\b{r}_1\b{r}_2;\b{r}_1'\b{r}_2';t)&=&\sum_{ijkl}\phi_i(\b{r}_1)\phi_j(\b{r}_2)g_{\tn{col},ijkl}^{\tn{ab},<}(t)\phi^*_k(\b{r}_1')\phi^*_l(\b{r}_2')\,,\\
g_{\tn{col},ijkl}^{\tn{ab},<}(t)&=&\i \int\limits_0^t \d \bar t \sum_{mnrs} V^\tn{ab}_{mnrs}
  \bigg\{g^{\tn{a},>}_{i,m}(t,\bar t)g^{\tn{b},>}_{j,n}(t,\bar t)\times \nonumber\\
	&&\left[ g^{\tn{a},<}_{k,r}(\bar t, t)g^{\tn{b},<}_{l,s}(\bar t, t)
    \pm \delta_\tn{ab}\, g^{\tn{a},<}_{l,r} (\bar t, t)g^{\tn{b},<}_{k,s} (\bar t, t) \right] \\ 
    &&-(> \leftrightarrow < \textnormal{and}\; g_{1,2} \leftrightarrow g_{2,1})
  \bigg\}\nonumber\,.
\end{eqnarray}
Basis expansions for inhomogeneous quantum many-body systems have been successfully applied to electrons in quantum dots (``artificial atoms''), e.g., \cite{balzer09}, and small atoms and molecules, e.g., \cite{dahlen07,balzer10}. For these systems, the Keldysh/Kadanoff-Baym equations are solved for the matrix function $g_{ij}^<(t,t')$. Using these results and formulas (\ref{eq:gablessHFtbasis}),(\ref{eq:gablessICtbasis}) and (\ref{eq:gablesscoltbasis}), the two-particle correlation function, $g^{\tn{ab},<}$ in configuration space, \refeqn{eq:gablessbasis}, can be reconstructed. Besides, also the matrix elements $g_{ijkl}^{\tn{ab},<}(t)$ themselves are of interest, as they carry extensive information on the many-body system. For example, the matrix components $g_{ijij}^{\tn{ab},<}(t)$ describe the correlation of two particles ``a,b'' occupying the orbitals i and j, respectively, at a given moment $t$. 
%-----------------------------
\section{Equilibrium pair distribution function}
\label{sec:eqpdf}
To compute the pair distribution in thermodynamic equilibrium from the Matsubara Green's function we return to \refeqn{eq:bse_static_born_short_1t} for the Keldysh matrix and extract the Matsubara component. According to \refeqn{eq:g0abalpha} it is given by a product of one-particle Matsubara Green's functions,
\begin{equation}
\label{eq:g0abM}
G_0^\tn{ab,M}(\b{r}_1\b{r}_2;\b{r}_1'\b{r}_2';\tau)=g^\tn{a,M}(\b{r}_1\b{r}_1',\tau)g^\tn{b,M}(\b{r}_2\b{r}_2',\tau)\,,
\end{equation}
where $\tau=t-t'$. Similarly as in nonequilibrium, the two-particle Matsubara Green's function consists of a Hartree-Fock and correlation part,
\begin{equation}
\label{eq:gabM}
g^\tn{ab,M}(\tau)=g^\tn{ab,M}_\tn{HF}(\tau)+g^\tn{ab,M}_\tn{col}(\tau)\,
\end{equation}
where the latter is obtained from \refeqn{eq:g0abalpha} using the Langreth rules, in analogy to \refeqn{eq:gablessIC},
\begin{equation}
g^\tn{ab,M}_\tn{col}(\tau)=-\i\int_0^\beta\d\bar{\tau}\,G_0^\tn{ab,M}(\tau-\bar{\tau})\Sigma^\tn{ab,M}_0(\bar{\tau})\,.
\end{equation}
Restoring the coordinate arguments and using the definitions of $G_0^\tn{ab}$ and $\Sigma_0^\tn{ab}$, Eqs.~(\ref{eq:G0ab}) and (\ref{eq:Sigma0ab}), we obtain
\begin{eqnarray}
\label{eq:gabMHF}
g^{\tn{ab,M}}_\tn{HF}({\bf r}_1{\bf r}_2;{\bf r}'_1{\bf r}'_2)&=&-\rho^\tn{a}({\bf r}_1{\bf r}'_1)\rho^\tn{b}({\bf r}_2{\bf r}'_2) \mp \delta_\tn{ab} \rho^\tn{a}({\bf r}_1{\bf r}'_2)\rho^\tn{b}({\bf r}_2{\bf r}'_1)\,,\\
g^{\tn{ab,M}}_\tn{col}({\bf r}_1{\bf r}_2;{\bf r}'_1{\bf r}'_2,\tau)&=&-\i\int_0^\beta \d\bar \tau \int \d^3{\bar r}_1 \d^3{\bar r}_2 V^\tn{ab}({\bar r}_{12})\times
\nonumber\\
&&\bigg\{
g^{\tn{a,M}}({\bf r}_1{\bar {\bf r}}_1,{\tau-\bar \tau}) g^{\tn{b,M}}({\bf r}_2{\bar {\bf r}}_2,{\tau-\bar \tau}) \times
\\\nonumber
&&\big[
g^{\tn{a,M}}({\bar {\bf r}}_1{\bf r}'_1,\bar \tau) g^{\tn{b,M}}({\bar {\bf r}}_2{\bf r}'_2,\bar \tau)
\pm \delta_\tn{ab}
g^{\tn{a,M}}({\bar {\bf r}}_1{\bf r}'_2,\bar \tau) g^{\tn{b,M}}({\bar {\bf r}}_2{\bf r}'_1,\bar \tau)
\big]\nonumber\bigg\}\,.\nonumber
\label{eq:gabMcol}
\end{eqnarray}
From this the equilibrium PDF is obtained by introducing center of mass and relative coordinates as was done in the nonequilibrium situation, cf. \refsec{sec:NeqPDFcoord}.
%
%-----------------------------
\section{Numerical example: pair correlations in an electron-hole bilayer system}
\label{sec:bilayer}
To illustrate the results obtained so far we consider an example of a two-component system where strong correlations play a prominent role. The system of interest consists of two layers of zero thickness which contain an equal finite number $N$ of negative (electrons) and positive (holes) charged particles with a finite layer spacing $d$. In each plane, the particles are confined by a harmonic potential of frequency $\Omega$. Such electron-hole bilayers have been actively studied in recent years, e.g., \cite{depalo02,filinov03,hartmann05}, because they allow to study strongly correlated excitons which may exhibit Bose condensation and superfluidity, as well as liquid-like and crystal-like behavior. For more details, we refer to Refs.~\cite{ludwig07,boening11}.
\subsection{Model}
The Hamiltonian of the quasi-two-dimensional electron-hole bilayer is given by
\begin{eqnarray}
\label{eq:H}
H&=&H_\tn{e}+H_\tn{h}+H_\tn{eh} \\
H_\tn{e}&=&\sum_{i=1}^{N_\tn{e}}\frac{1}{2}\left(-\Delta_{i,\e} +{\bf r}_{i,\e}^2\right)+
 \lambda\sum_{i<j=2}^{N_\e}\frac{1}{\sqrt{({\bf r}_{i,\e}-{\bf r}_{j,\e})^2}}\,,\\
 H_\tn{h}&=&\sum_{i=1}^{N_\tn{h}}\frac{1}{2}\left(-\frac{m^*_\e}{m^*_\tn{h}}\Delta_{i,\tn{h}}+
  \frac{m^*_\tn{h}}{m^*_\e}{\bf r}_{i,\tn{h}}^2\right)+
\lambda\sum_{i<j=2}^{N_\tn{h}}\frac{1}{\sqrt{({\bf r}_{i,\tn{h}}-{\bf r}_{j,\tn{h}})^2}}\,,\\
H_\tn{eh}&=&-\lambda\sum_{i=1}^{N_\e}\sum_{j=1}^{N_\tn{h}}\frac{1}{
  \sqrt{({\bf r}_{i,\e}-{\bf r}_{j,\tn{h}})^2+d^{*2} } }\,,
\end{eqnarray} 
where we introduced dimensionless variables by rescaling length and energy by the units,
\begin{equation}
r_0=\sqrt{\frac{\hbar}{m_\e\Omega}}\,,\qquad E_0=\hbar\Omega\,.
\end{equation}
The term $m_\tn{e/h}^*$ denotes the effective mass of electrons/holes and $d^*$ is the effective distance between the layers.
Further, we introduced the coupling parameter $\lambda$ measuring the strength of the Coulomb interaction relative to the confinement energy
\begin{equation}
\lambda=\frac{r_0}{a_\tn{B}}\,,\qquad a_\tn{B}=\frac{\hbar^2}{m_\e e^2}\,.
\end{equation}
Here $a_\tn{B}$ is the Bohr radius of a Hydrogen-like bound state---an exciton (we use the electron mass instead of the reduced mass). 

The coupling parameter measures the strength of the Coulomb interaction among identical particles in each layer as well as the correlation between electron and holes. Generally, one may expect that for $\lambda \leq 1$, i.e. for very strong confinement, there is a strong wave function overlap, and the system approaches ideal gas like behavior. In the opposite case, $\lambda \gg 1$, the Coulomb interaction dominates, and particles will tend to become localized. Finally, variation of the layer separation gives an additional control of the many-particle state: for $d\ra \infty$ both layers will be decoupled, containing independent electron and hole populations, whereas for decreasing $d$ Coulomb attraction plays an increasing role. This gives rise to formation of indirect excitons which behave (approximately) as bosons and exhibit dipole interaction, e.g., \cite{ludwig07,boening11}.
%---
\subsection{Equilibrium PDF of the electron-hole bilayer---Comparison to Path Integral Monte-Carlo results}
\begin{figure}
\label{fig:XPDF}
\includegraphics[width=16cm]{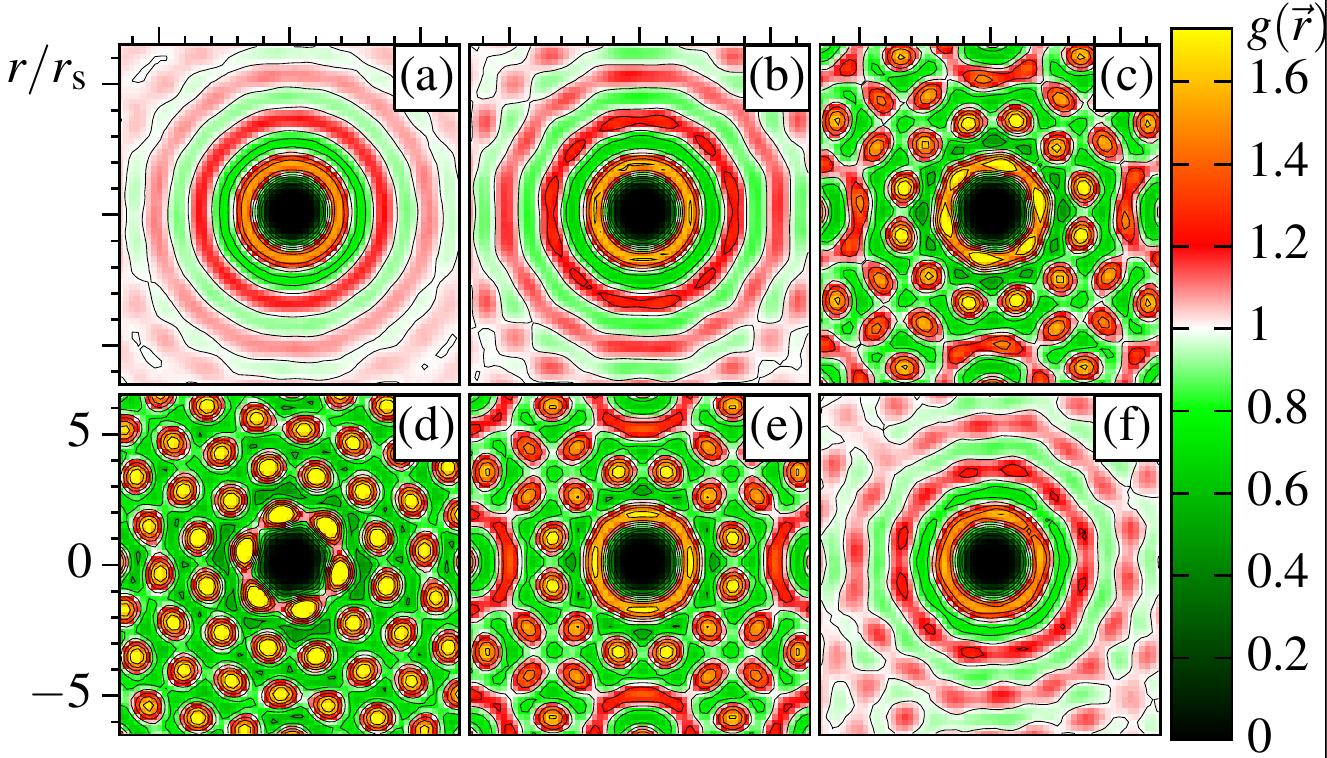}
\caption{Radial Exciton-Exciton PDF $h^{\tn{X}_c\tn{X}}(\b{r})$ in a ZnS$_{0.3}$Se$_{0.7}$/ZnSe quantum well at $T=0.63\,\tn{K}$ with layer separation $d=40.83\,\tn{nm}$. Densities in units of $1.06\cdot 10^{17}\tn{m}^{-2}$: (a) $0.84\cdot 10^{-3}$, (b) $1.3\cdot 10^{-3}$, (c) $1.7\cdot 10^{-3}$, (d) $3.2\cdot 10^{-3}$, (e) $3.6\cdot 10^{-3}$, (f) $4.0\cdot 10^{-3}$. The radial length is measured in units of the so called Brueckner parameter $r_\tn{s}=3.26\cdot 10^8\times a$, where $a$ is the mean interparticle distance. The illustration is taken from Ref.~\cite{boening11}.}
\end{figure}

%---
Preliminary results for the equilibrium PDF of the electron-hole bilayer according to Sec.~\ref{sec:eqpdf} were obtained recently \cite{kobusch12}, however they still require further numerical tests. Therefore, to illustrate the physical content of the PDF, in this section, we show some results obtained from path integral Monte-Carlo (PIMC) calculations by \textit{Böning~et~al.} \cite{boening11}. Their approach to the description of the bilayer system is different to the one presented in \refsec{sec:bilayer}, as they assume beforehand, that the layer separation $d$ and the interaction strength $\lambda$ in \refeqn{eq:H} between the electrons and holes, respectively, induces the formation of indirect excitons, i.e. quasiparticles, comprised of strongly bound but spatially separated electron-hole pairs. The excitons exhibit an interaction, which for large distances is of dipole type and for small distances approaches a soft Coulomb potential, for details see Ref.~\cite{boening11}. With these assumptions \textit{Böning~et~al.} computed the equilibrium PDF $h^{\tn{X}_c\tn{X}}(\b{r})$ of one exciton $X$ relative to a fixed exciton $X_c$ in the center for different values of the exciton density $n$ in a ZnS$_x$Se$_{1-x}$/ZnSe quantum well with doping factor $x=0.3$. The density is measured in units of $a_\tn{B}^{*-2}=1.06\cdot 10^{17}\tn{m}^{-2}$, where $a_\tn{B}^{*}=\hbar^2\epsilon/(e^2m_e^*)$ is the electronic Bohr radius with the material constants $\epsilon=8.7$ and $m_e^*=0.15\,m_0$. The temperature is chosen to be $k_\tn{B}T=0.001\,\tn{Ha}^*$, where the energy unit is defined as $\tn{Ha}^*=e^2/(\epsilon a_\tn{B}^*)=53.93\,\tn{meV}$, resulting in a temperature of $T=0.63\,\tn{K}$. In \reffig{fig:XPDF}, $h^{\tn{X}_c\tn{X}}(\b{r})$ is shown for different densities.
One can see that for low densities, the excitons in the system are melted, showing no localization or radial ordering. If the density is increased, an exciton crystal starts to form, due to pressure crystallization. For further increased density this exciton crystal melts again, which is an effect of the Coulomb-like character of the exciton-exciton interaction at short distances, the excitons undergo quantum melting. 
The advantage of the PDF is clear from this figure: different phases of the system can be clearly distinguished which is not possible on the basis of single-particle quantities such as the density. 

PIMC is very efficient for computing the thermodynamic properties and also spectral properties \cite{filinov12} of bosons, such as excitons. However, at high densities, excitons break up and form an electron-hole plasma (Mott effect) consisting of fermions. PIMC simulations of fermions at low temperatures are still hampered by the notorious sign problem \cite{schoof11}. In contrast, with nonequilibrium Green's functions, this regime is easily accessible whereas limitations arise with increasing coupling strength. Therefore, PIMC and NEGF have complementary areas of applicability. Moreover, NEGF should allow one to access the time-evolution of the PDF, as demonstrated above.          
%
%-----------------------------
\section{Discussion}
\label{sec:discussion}
In this paper we presented an approach to the pair distribution function of a quantum many-body system in the frame of the nonequilibrium real-time Green's functions. This problem is complicated due to the fact that the standard approach used in NEGFs uses a formal decoupling of the Martin-Schwinger hierarchy on the level of the first equation: the two-particle Green's function is eliminated by introduction of the single-particle selfenergy, cf. \refeqn{eq:Selfenergy}. With this elimination also direct access to the pair correlations is lost. Thus one has to reconstruct the pair correlations form the single-particle Green's function within a chosen approximation for the selfenergy. 

To solve this reconstruction problem we started the analysis from the equation of motion of the two-particle Green's function---the Bethe-Salpeter equation and simplified it systematically. We concentrated on the case of the static second Born approximation because for it a large number of numerical solutions of the KBE exist, for which it would be desirable to evaluate the PDF. It was shown that in second Born approximation a closed expression for the nonequilibrium PDF can be derived which can be straightforwardly evaluated. The result involves combinations of four single-particle Green's functions and is computationally expensive. Numerical results will be presented in a forthcoming paper.   

\ack
We thank Th. Bornath for stimulating discussions. This work is supported in part by the Deutsche Forschungsgemeinschaft via SFB-TR24 and project BO1366/9.

%\appendix*\section{}

\section*{References}
\begin{thebibliography}{12}
\bibitem{danielewicz84} Danielewicz~P 1984 {\it Ann.~Phys.} (N.Y.) \b{152} 304
\bibitem{garny09} Garny~M and Müller~M~M 2009 {\it Phys.~Rev.} D \b{80} 085001
\bibitem{haug96} Haug~H and Jauho~A-P 1996 {\it Quantum Kinetics in Transport and Optics of Semiconductors}, Springer 
\bibitem{bonitz12} Bonitz~M, Henning~C and Block~D 2010 {\it Reports~Prog.~Physics} \b{73} 066501
\bibitem{kremp05} Kremp~D, Schlanges~M and Kraeft~W-D 2005 {\it Quantum Statistics of Nonideal Plasmas}, Springer
\bibitem{krausz09} Krausz~F and Ivanov~M 2009 {\it Rev.~Mod.~Phys.} {\bf 81} 163
\bibitem{krasovskii07} Krasovskii~E~E and Bonitz~M 2007 {\it Phys.~Rev.~Lett.} {\bf 99} 247601
\bibitem{becker08} Becker~U 2008 {\it Nature~Physics} {\bf 5} 649
\bibitem{bonitz96} Bonitz~M and Kremp~D 1996 {\it Phys.~Lett.} A \b{212} 83
\bibitem{kremp97} Kremp~D, Bonitz~M, Kraeft~W~D and Schlanges~M 1997 {\it Ann.~Phys.} (N.Y.) \b{258} 320
\bibitem{bonitz98} Bonitz~M 1998 {\it Quantum Kinetic Theory}, Teubner, Stuttgart, Leipzig
\bibitem{bauch08} Bauch~S 2008 {\it Phys.~Rev.} A \b{78} 043403 
\bibitem{bonitz96j} Bonitz~M, Kremp~D, Scott~D~C, Binder~R, Kraeft~W~D and Köhler~H~S 1996 {\it J.~Phys.~Cond.~Matt.} \b{8} 6057
\bibitem{bonitz97} Bonitz~M, Semkat~D and Kremp~D 1997 {\it Phys.~Rev.} E \b{56} 1246
\bibitem{kwong00} Kwong~N~H and Bonitz~M 2000 {\it Phys.~Rev.~Lett.} \b{84} 1768
\bibitem{dahlen07} Dahlen~N~E and van~Leeuwen~R 2007 {\it Phys.~Rev.~Lett.} \b{98} 153004
\bibitem{stan09} Stan~A, Dahlen~N~E and van~Leeuwen~R 2009 {\it J.~Chem.~Phys.} \b{103} 176404
\bibitem{vonfriesen09} von~Friesen~P, Verdozzi~C and Almbladh~C-O 2009 {\it Phys.~Rev.~Lett.} \b{103} 176404
\bibitem{vonfriesen11} von~Friesen~P, Verdozzi~C and Almbladh~C-O 2011 {\it Europhys.~Lett.} \b{95} 27005
\bibitem{bonitz96p} Bonitz~M 1996 {\it Phys.~Lett.} A \b{221} 85
\bibitem{depalo02} De~Palo~S, Rapisarda~F and Senatore~G 2002 {\it Phys.~Rev.~Lett.} \b{88} 206401
\bibitem{filinov03} Filinov~A, Ludwig~P, Golubnychiy~V, Bonitz~M and Lozovik~Yu~E 2003 {\it phys.~stat.~sol.} (c) \b{0} 1518
\bibitem{hartmann05} Hartmann~P, Donko~Z and Kalman~G 2005 {\it Europhys.~Lett.} \b{72} 396
\bibitem{schleede12} Schleede~J, Filinov~A, Bonitz~M and Fehske~H 2012 {\it Contrib.~Plasma~Phys.} DOI: 10.1002/ctpp.201200045 
\bibitem{filinov09} Filinov~A, Ludwig~P, Bonitz~M and Lozovik~Yu~E 2009 {\it J.~Phys.} A \b{42} 214016
\bibitem{ludwig07} Ludwig~P, Filinov~A, Lozovik~Yu~E, Stolz~H and Bonitz~M 2007 {\it Contrib.~Plasma~Phys.} \b{47} 335
\bibitem{ludwig08} Ludwig~P, Balzer~K, Filinov~A, Stolz~H and Bonitz~M 2008 {\it New~Journal~of~Physics} \b{10} 083031 
\bibitem{kalman07} Kalman~G, Hartmann~P, Donko~Z and Golden~K 2007 {\it Phys.~Rev.~Lett.} \b{98} 236801
\bibitem{boening11} Böning~J, Filinov~A and Bonitz~M 2011 {\it Phys.~Rev.} B \b{84} 075130
\bibitem{boercker79} Boercker~D~B and Dufty~J~W 1979 {\it Ann.~Phys.} (N.Y.) \b{119} 43
\bibitem{semkat99} Semkat~D, Kremp~D and Bonitz~M 1999 {\it Phys.~Rev.} E \b{59} 1557
\bibitem{semkat00} Semkat~D, Kremp~D and Bonitz~M 2000 {\it J.~Math.~Phys.} \b{42} 7458
\bibitem{semkat03} Semkat~D, Bonitz~M and Kremp~D 2000 {\it Contrib.~Plasma~Phys.} \b{43} 321 
\bibitem{bornath99} Bornath~T, Kremp~D and Schlanges~M 1999 {\it Phys.~Rev.} B \b{60} 6382
\bibitem{balzer09} Balzer~K, Bonitz~M, van~Leeuwen~R, Dahlen~N~E and Stan~A 2009 {\it Phys.~Rev} B \b{79} 245306
\bibitem{balzer10} Balzer~K, Bauch~S and Bonitz~M 2010 {\it Phys.~Rev} A \b{81} 022510
\bibitem{lipavsky86} Lipavski~P, Spicka~V and Velicky~B 1986 {\it Phys.~Rev.} B \b{34} 6933
\bibitem{hermanns12} Hermanns~S, Balzer~K and Bonitz~M 2012 {\it J.~Phys.~Conf.~Series} this issue 
\bibitem{rosenthal09} Rosenthal~L 2009 {\it Green's functions approach to electron-hole bilayers} (in German), Diploma thesis, Kiel University
\bibitem{kobusch12} Kobusch~K 2012 {\it Nonequilibrium Green's function approach to the pair distribution function in the second Born approximation} (in German), Diploma thesis, Kiel University
\bibitem{filinov12} Filinov~A, Prokof'ev~N~V and Bonitz~M 2010 {\it Phys.~Rev.~Lett.} \b{105} 070401, \\
Filinov~A and Bonitz~M 2012 {\it Phys.~Rev.} A, in press, arXiv:1205.5191
\bibitem{schoof11} Schoof~T, Bonitz~M, Filinov~A, Hochstuhl~D and Dufty~J~W 2011 {\it Contrib.~Plasma~Phys.} \b{51} 687-697   
\end{thebibliography}
\end{document}